\documentclass{jetplfree}
\twocolumn
\usepackage[T2A]{fontenc}
\usepackage[cp1251]{inputenc}
\usepackage[english]{babel}
\usepackage{graphicx}
\usepackage{dblfloatfix}
\usepackage{fixltx2e}
\usepackage{amsmath}
\usepackage{mathtools}
\usepackage[hyphens]{url}
\usepackage{xcolor}
\usepackage[noadjust]{cite}

\title{A spatially resolved network spike in model neuronal cultures reveals nucleation centers, circular traveling waves and drifting spiral waves}

\rtitle{A spatially resolved network spike in model neuronal cultures reveals nucleation centers...}

\author{A.\,V.\,Paraskevov$^{1,2}$, D.\,K.\,Zendrikov$^{2,1}$}

\rauthor{A.\,V.\,Paraskevov, D.\,K.\,Zendrikov}

\sodauthor{A.\,V.\,Paraskevov, D.\,K.\,Zendrikov}

\address{$^1$National Research Centre "Kurchatov Institute", 123182 Moscow, Russia,\\~\\
$^2$Moscow Institute of Physics and Technology (State University), 141700 Dolgoprudny, Russia}

\abstract{
We show that in model neuronal cultures, where the probability of interneuronal connection formation decreases exponentially with increasing distance between the neurons, there exists a small number of spatial nucleation centers of a network spike, from where the synchronous spiking activity starts propagating in the network typically in the form of circular traveling waves. The number of nucleation centers and their spatial locations are unique and unchanged for a given realization of neuronal network but are different for different networks. In contrast, if the probability of interneuronal connection formation is independent of the distance between neurons, then the nucleation centers do not arise and the synchronization of spiking activity during a network spike occurs spatially uniform throughout the network. Therefore one can conclude that spatial proximity of connections between neurons is important for the formation of nucleation centers. It is also shown that fluctuations of the spatial density of neurons at their random homogeneous distribution typical for the experiments \textit{in vitro} do not determine the locations of the nucleation centers. The simulation results are qualitatively consistent with the experimental observations.

\bigskip

Keywords: cultured neuronal network, synaptic plasticity, network spike, spatial dynamics, nucleation centers, circular traveling wave, multiarmed spiral wave
}

\begin{document}

\maketitle

\textbf{1. Introduction}

\bigskip

In neuronal cultures, i.e. planar neuronal networks grown \textit{in vitro} from initially dissociated neurons typically of cerebral cortex or hippocampus, one can often detect spontaneous short-term (fractions of a second) repetitive synchronization of neuronal spiking activity called a network spike or a population burst \cite{1,2,3,4,5}. This phenomenon is thought to be related to epilepsy \cite{6,7} therefore both the origin and the properties of network spikes are the subject of intensive studies \cite{8,9,10,11,12,13,14,15,16,17,18,19}. The particular attention is paid to identifying patterns of the network spike initiation \cite{2,3,4,20,21,22,23,24,25,26,27}. In a recent paper \cite{4} (see also \cite{20,28,29,30}) it has been shown experimentally that a typical network spike has a few steady spatial sources - nucleation centers of traveling waves of synchronous spiking activity. The causes of their occurrence have not yet been identified. As described in \cite{4}, the number and locations of the nucleation centers for different neuronal cultures are different, but for the same neuronal culture these remain practically unchanged during the observation period.

In this paper, by means of simulations, we investigated the spatial dynamics of network spikes in large planar neuronal networks (50 thousand neurons, several millions of interneuronal connections), which are comparable to real neuronal cultures. It was suggested that the probability $p_{con}(r)$ of an unidirectional connection between two neurons decreases exponentially as a function of the distance $r$ between them \cite{31}. In fact, we have generalized the model \cite{8}, where the network spikes occurred in a neuronal network composed of Leaky Integrate-and-Fire (LIF) neurons with binomial distribution of interneuronal connections and relaxational synaptic plasticity, for the case of spatially dependent network topology, taking into account the respective propagation delays of signals between neurons.

We have found that (i) for the network of excitatory neurons, uniformly distributed over the square area, there is indeed a small number of spontaneously-formed nucleation centers of a network spike, from where the synchronous spiking activity propagates farther typically in the form of circular traveling waves. The number of nucleation centers and their spatial locations are unique and invariable for a given implementation of the neuronal network, but are different for different networks. The nucleation centers are not nested in fluctuations of spatial density of neurons and the changes in the function $p_{con}(r)$, under certain conditions on the average values of network parameters, do not lead to the disappearance of nucleation centers. (ii) If the probability of formation of interneuronal connection is not dependent on the neurons' location relative to each other, then the nucleation centers do not arise - the synchronization of spiking activity occurs spatially uniform throughout the network. (iii) In the networks of excitatory and inhibitory neurons with relatively high density of interneuronal connections, a network spike may occasionally occur in the form of non-stationary multiarmed spiral wave with the drifting center.

The results obtained, in particular, the existence of nucleation centers and the statistical regularities of their occurrence, seem consistent with the spatial dynamics of network spikes described in \cite{4}. It is worth noting that the dynamic transitions between the phases of asynchronous and synchronous spiking activity of the network could be related to phase transitions of either the first (case (i)) or the second (case (ii)) kind, depending on the degree of locality of the majority of interneuronal connections.

\bigskip

\textbf{2. Neuronal network model}

\bigskip

A mathematical model of the neuronal network comprised of three main components: (I) the model of a neuron, (II) synapse model describing the interaction between neurons, and (III) algorithm for generating the network topology. By default, the network consisted of 80\% excitatory and 20\% inhibitory neurons. The values of parameters for the neuron and synapse models, including the parameters of normal distributions (standard deviations of which by default were taken equal to 1/2 of the average values), do not differ essentially from those used in article \cite{8} (see \cite{32}).

I. As a neuron model, the standard LIF-neuron has been used. Subthreshold dynamics of the transmembrane potential $V$ of such a neuron is described by the equation
\begin{equation}
\tau _{m}dV/dt=V_{rest}-V(t)+(I_{syn}(t)+I_{bg})R_{m}, \label{LIF}
\end{equation}
where $V_{rest}$ is the neuron's resting potential, $\tau_{m}$ is the characteristic time for relaxation of $V$ to $V_{rest}$, $R_{m}$ is the electrical resistance of the neuron's membrane, $I_{syn}(t)$ is the total incoming synaptic current, which, as a function of time $t$, depends on the choice of the dynamic model of a synapse and the number of incoming synapses, $I_{bg}$ is a constant "background" current, the magnitude of which varies from neuron to neuron by a normal distribution. The background currents are required in order to initiate and sustain a spontaneous asynchronous spiking activity of the network. These also determine the diversity of neuronal excitability in the network.

When the transmembrane potential reaches a threshold value $V_{th}=V(t_{sp})$, it is supposed that the neuron emits a spike, then $V$ abruptly drops to a specified value $V_{reset}$, $V_{rest}<V_{reset}<V_{th}$, and retains this value during the period of refractoriness $\tau_{ref}$, then the dynamics of the potential is again described by the equation \eqref{LIF}. The result of the LIF-neuron dynamics is a sequence of spike generation moments $\{t_{sp}^{(1)}, t_{sp}^{(2)},\ldots\}$.

If a neuron has the value of $I_{bg}$ that exceeds a critical value $I_{c}=(V_{th}-V_{rest})/R_{m}$, then this neuron is a pacemaker, i.e., it is able to emit spikes periodically, with the period $\triangle t_{sp}=\tau_{ref}+\tau_{m}\ln[(I_{bg}-I_{r})/(I_{bg}-I_{c})]$, where $I_{r}=(V_{reset}-V_{rest})/R_{m}$, in the absence of incoming signals from other neurons. Our network model implies that both excitatory and inhibitory neurons may be pacemakers.

Finally, it is worth noting that the LIF-neuron has no ability for intrinsic bursting, unlike the neuron model used in \cite{4,17}.

II. A single contribution to the incoming synaptic current in the TUM model \cite{8} is determined by the formula
\begin{equation}
I_{syn}(t)=A\cdot y(t), \label{Isyn}
\end{equation}
where $A$ is the maximum amplitude of synaptic current, the sign and magnitude of which depend on the type of pre- and postsynaptic neurons (i.e., whether the neuron is excitatory or inhibitory), and $y(t)$ is a dimensionless parameter, $0\le y\le 1$, the dynamics of which is determined by the following system of equations:
\begin{equation}
\left\{\begin{array}{l} {dx/dt=z/\tau_{rec} -u\cdot x\cdot \delta (t-t_{sp}-\tau_{del})}, \\
{dy/dt=-y/\tau_{I} +u\cdot x\cdot \delta (t-t_{sp}-\tau_{del})}, \\
{dz/dt=y/\tau_{I} -z/\tau_{rec}}, \end{array}\right. \label{xyz}
\end{equation}
where $x$, $y$, and $z$ are the fractions of synaptic resources in the recovered, active and inactive state, respectively, $x+y+z=1$, $\tau_{rec}$, $\tau_{I}$ are the characteristic relaxation times, $\delta(\ldots)$ is the Dirac delta function, $t_{sp}$ is the moment of spike generation at the presynaptic neuron, $\tau_{del}$ is the spike propagation delay (see \eqref{tau_del}), and $u$ is the fraction of recovered synaptic resource used to transmit the signal across the synapse, $0\le u\le 1$. For the outgoing synapses of inhibitory neurons, the dynamics of $u$ is described by the equation
\begin{equation}
du/dt=-u/\tau _{facil} +U\cdot (1-u)\cdot \delta (t-t_{sp}-\tau_{del}),  \label{u}
\end{equation}
where $\tau_{facil}$ is the characteristic relaxation time, and $0<U\le 1$ is a constant parameter. For the outgoing synapses of excitatory neurons, $u$ remains constant and equals to $U$. In the numerical simulations the constants $A$ and $U$, as well as all the characteristic relaxation times (except for $\tau_{I}$) in the synaptic current model, were normally distributed, i.e. each synapse had its own unique values of these parameters.

III. We used the binomial and spatially-dependent distributions of interneuronal connections. To simplify the model, the formation of autaptic connections was prohibited. In the case of "binomial" network topology, we set a constant probability $p_{con}$ of the formation of unilateral synaptic connection between two neurons, independent of their spatial coordinates. Then in the network of $N$ neurons the number $m$ of outgoing connections of a neuron is described by the binomial distribution $P(m)=C_{N-1}^{m} p_{con}^{m}(1-p_{con})^{N-1-m}$, where $0\leq m\leq N-1$, with mean value $\bar{m}=p_{con}(N-1)$.

In the case of spatially-dependent network topology, point neurons were uniformly distributed over a square area $L\times L$ of unit size ($L=1$). The probability of formation of unilateral connection between each pair of neurons depended on the distance $r$ between them according to the formula \cite{31}
\begin{equation}
p_{con}(r)=Be^{-r/\lambda}, \label{p_con}
\end{equation}
where $\lambda$ is the characteristic connection length, expressed in units of $L$. The constants $B$ and $\lambda$, for simplicity, were chosen independent of the types of pre- and postsynaptic neurons, in particular, it was taken $B=1$, $\lambda = const$ for all combinations of types of neurons.

Note two essential circumstances: first, since the square area is a convex set of points, we assumed that the interneuronal connections may be modeled by segments of straight lines. In addition, as the connections do not cross boundaries of the square, the neurons in the vicinity of the boundaries have fewer connections. Secondly, despite the fact that $p_{con}(r)$ reaches its maximum at $r=0$, the distribution of the lengths of interneuronal connections is zero at $r=0$ and reaches its maximum at the point $r\approx \lambda$, provided that $\lambda\lesssim0.1$. One can show this straightforwardly by finding the probability density $P(r)$ to detect two neurons at a distance $r$ from each other,
\begin{equation}
P(r)=
\begin{dcases}
2r\cdot(\pi-4r+r^{2}), \text{ } r\le 1, \\
4r\cdot(2\arcsin(1/r)+2\sqrt{r^{2}-1}-\dots \\-\pi/2-r^{2}/2-1), \text{ } 1<r\le\sqrt{2},
\end{dcases}\label{P_of_r}
\end{equation}
such that ${\displaystyle\int\limits_{0}^{\sqrt{2}}}P(r)dr=1$ \cite{32,33,34,35}. The distribution of interneuronal connection lengths is given by the product $p_{con}(r)P(r)$ (Fig. 1, upper graph), cp. \cite{4,17,36}. In turn, the average number of interneuronal connections in the network of $N$ neurons is
\begin{equation}
N_{con}(\lambda)=N(N-1){\int\limits_{0}^{\sqrt{2}}}p_{con}(r)P(r)dr, \label{N_con}
\end{equation}
so that the corresponding probability for the binomial distribution can be found as $p_{con}=N_{con}(\lambda)/(N(N-1))$ (Fig. 1, lower graph). The approximate analytical expression for the function $N_{con}(\lambda)$ is given in \cite{32}.
\begin{figure}[!t]
\centering
\includegraphics[width=0.45\textwidth]{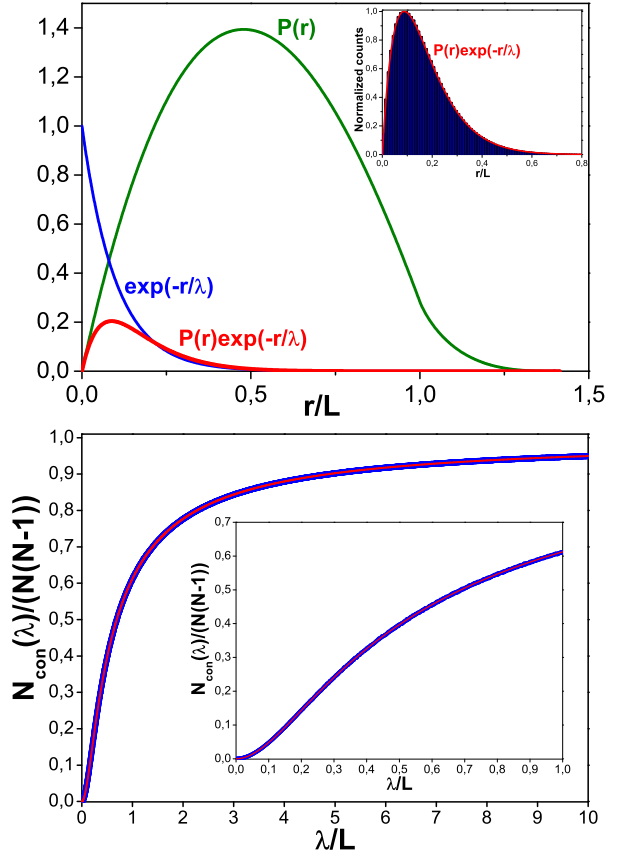} 	
\caption{Figure 1. Upper graph: Probability density $P(r)$ to find two neurons, whose spatial coordinates are at random uniformly distributed in the square $L\times L$, at distance $r$ from each other. Inset: Normalized distribution of the interneuronal connection lengths for the network of $N=10^4$ neurons at $\lambda = 0.1L$ (see \eqref{p_con}) and the corresponding product $p_{con}(r)P(r)$. Lower graph: Functional dependence $N_{con}(\lambda)$ obtained (i) by direct simulation of the networks of $N=10^4$ neurons (thick blue curve) and (ii) with the use of the approximate analytical expression of Eq. \eqref{N_con} (thin red curve, which is virtually superimposed on the blue curve).}
\label{Fig1}
\end{figure}

The delays resulting from the propagation of spikes along the axons were calculated by the formula \cite{37}
\begin{equation}
\tau_{del}=\tau_{del, min}+r/v_{sp}, \label{tau_del}
\end{equation}
where $\tau_{del}$ is the total propagation delay of a spike along the axon of length $r$, $\tau_{del, min}$ is the minimal axonal delay the same for all synapses, and $v_{sp}$ is the constant speed of spike propagation along the axon \cite{32}. Note that the distribution of axonal delays \eqref{tau_del} is also determined by the product $p_{con}(r)P(r)$.

\bigskip

\textbf{3. Results}

\bigskip

Article \cite{8} lists the parameter values for the TUM model at which the regime of repetitive network spikes occurs in simulations (Fig. 2). This regime ("TUM regime") is characterized by a large variability of intervals between subsequent network spikes for different realizations of the network, cp. \cite{5,12,16}. (Findings \cite{9,11} indicate that the realization-averaged distribution of \textit{increments} of these intervals may be approximated by the Levy distribution.) It is important to note that the TUM regime occurs only in a relatively narrow region of values of the average number $\bar{m}=p_{con}(N-1)$ of outgoing connections per neuron.
\begin{figure}[!t]
\centering
\includegraphics[width=0.45\textwidth]{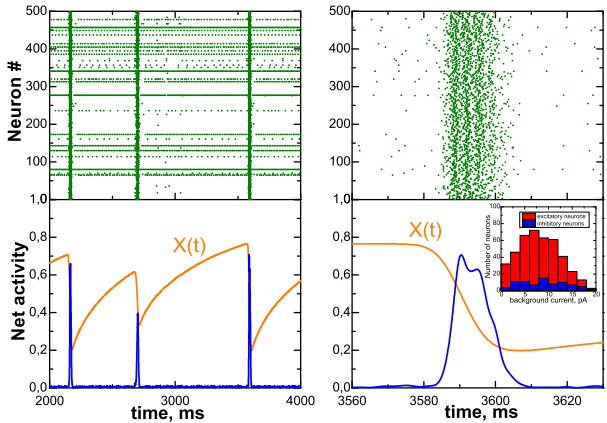} 	
\caption{Figure 2. Spiking activity of the "binomial" neuronal network of $N$ = 500 LIF-neurons (80\% excitatory, 20\% inhibitory; 6.4\% pacemaker neurons) with $p_{con}=0.1$ and the normal distribution of the background currents (see the inset at right). LEFT: Raster plot (top) and spiking activity, averaged over 2 ms and normalized to the total number of neurons (bottom). Network spikes are the vertical stripes in the raster and the peaks on the activity plot. $X(t)$ is the network-averaged fraction of synaptic resources in the recovered state. RIGHT: The same quantities for a single network spike.}
\label{Fig2}
\end{figure}
In particular, keeping unchanged other parameters of the simulations, for the networks of excitatory neurons the TUM-regime occurred in the range $30 \lesssim \bar{m} \lesssim 90$ (see Fig. 3). If the network comprises 20\% inhibitory neurons, this range is expanded, $30 \lesssim \bar{m} \lesssim 150$. In what follows, we obtain and examine the TUM regime for the case of spatially-dependent network topology. For planar networks with a large number of neurons (40-50 thousand) uniformly distributed over a square area, the parameter $\lambda$, which determines the probability \eqref{p_con} of interneuronal connection formation, was typically set so that the average number of outgoing connections per neuron ($\bar{m}=N_{con}(\lambda)/N$, see \eqref{N_con}) was inside this range, near its lower boundary for the sake of conserving computing resources.
\begin{figure}[!t]
\centering
\includegraphics[width=0.45\textwidth]{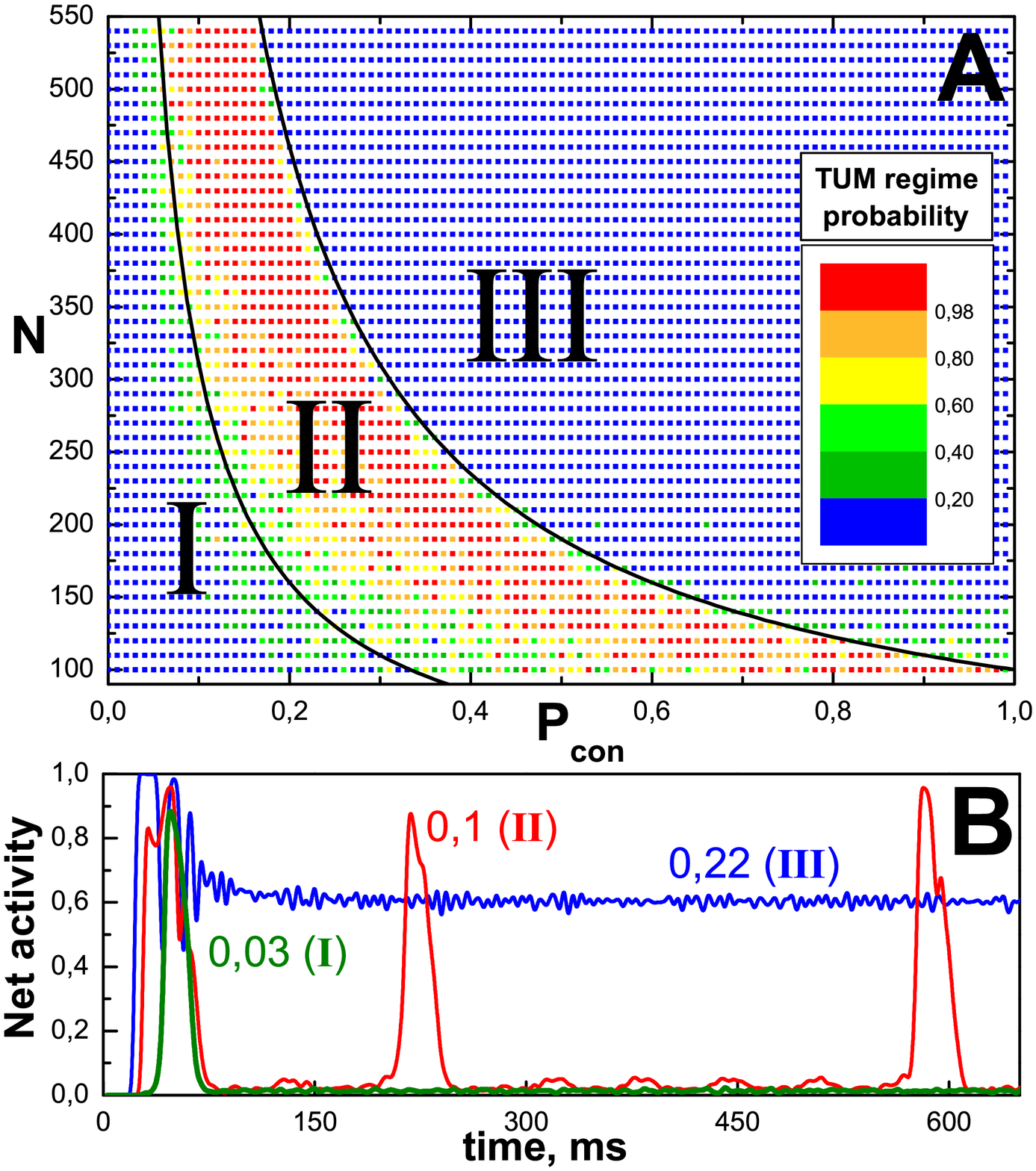}
\caption{Figure 3. \textbf{А}: Phase diagram of the occurrence probability for the regime of aperiodically repetitive network spikes (TUM-regime) in the "binomial" networks of excitatory neurons. $N$ is the number of neurons in the network, $p_{con}$ is the probability of interneuronal connection formation. Each pixel of the diagram displays the relative number of realizations of a neuronal network with given values of $N$ and $p_{con}$ (in total, there were 5 such trials) for which the TUM-regime was observed. \textbf{B}: Examples of averaged (over 3 ms) spiking activity of the network of $N$ = 500 excitatory neurons at $p_{con}$ = 0.03, 0.1, 0.22.}
\label{Fig3}
\end{figure}

In the TUM-regime, a network spike in the network of excitatory neurons uniformly distributed in the square area starts in one of a few (usually 3-4) spatial centers - primary nucleation centers, from which the synchronous spiking activity starts propagating through the network typically in the form of a circular traveling wave accompanied by the activation of more numerous secondary nucleation centers (Fig. 4). A spatial profile of the network spike emerging in the nucleation center is shown in Fig. 5.
\begin{figure*}[!t]
\centering
\includegraphics[width=0.75\textwidth]{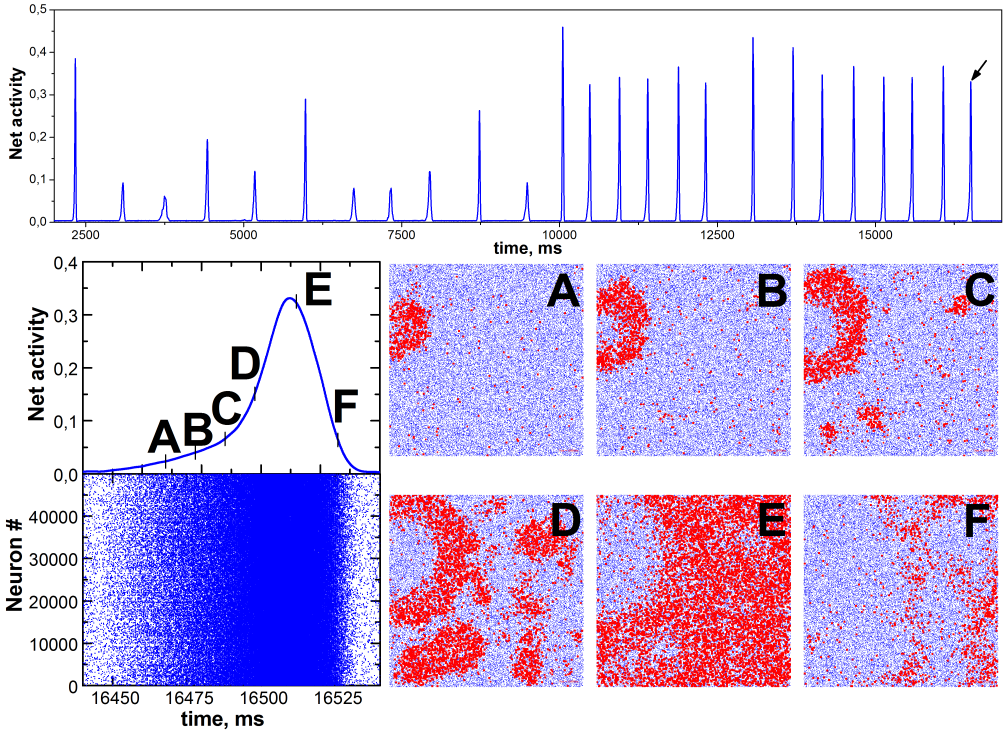}
\caption{Figure 4. Upper graph: Averaged (over 2 ms) spiking activity of the network of 50 thousand neurons at $\lambda = 0.01L$. After the first 10 seconds of the simulation the inhibitory neurons (20\% of the total) are blocked, i.e. do not participate in spiking dynamics of the network, in order to obtain a clear picture of the propagation of synchronous spiking activity from the smallest number of nucleation centers. Lower graph: LEFT: Network activity (top) and raster (bottom) during the network spike marked by the arrow in the upper graph. RIGHT: Snapshots of the instantaneous spatial activity of neurons for the corresponding moments of the network spike. Blue dots depict neurons and red dots highlight spiking neurons. Each frame corresponds to the area $L\times L$. On the frame \textbf{C} it is seen that in addition to the primary nucleation center (frame \textbf{A}) three secondary centers become active. The simulation parameters are described in detail in \cite{32}.}
\label{Fig4}
\end{figure*}

The primary nucleation centers are determined at the initial stage of a network spike by their invariable spatial arrangement (Fig. 5, right graph). The evaluation of their number, obviously, depends on the simulation time since the network spikes occur randomly in one of them with different relative probabilities. According to our observations (in total, 12 simulations of the same type and 14 various modifications were performed), the number of primary nucleation centers ceases to increase after 10-15 sequentially passed network spikes. We therefore conclude that it remains the same for a given realization of the neuronal network, being different for different networks. Note that a typical rate of generating network spikes is about a few hertz (Fig. 4) and the corresponding timescale has the same order of magnitude as the resource recovery time $\tau_{rec}$ (see \eqref{xyz}) for an outgoing synapse of an excitatory neuron \cite{32}. However, for some network realizations this similarity in the timescales is strongly violated.

Inhibitory neurons, in their turn, generally (i) decrease the average frequency of network spike occurrence, (ii) increase the variability of both the amplitude and duration of a network spike, and (iii) hinder the activity of primary nucleation centers and increase the number of secondary ones.

Interestingly, if the average number of outgoing synaptic connections per neuron is sufficiently large (i.e., parameter $\lambda$ in Eq. \eqref{p_con} is relatively large), then a multiarmed spiral wave with the drifting center can arise during some network spikes (Fig. 6), given that most of the network spikes still start with circular traveling waves diverging from the motionless nucleation centers. (In total, three such spirals occurred in two of five identical simulations at $\lambda$ = 0.04 with relative rates 1/12 and 2/10, respectively.)
\begin{figure*}[!t]
\centering
\includegraphics[width=0.75\textwidth]{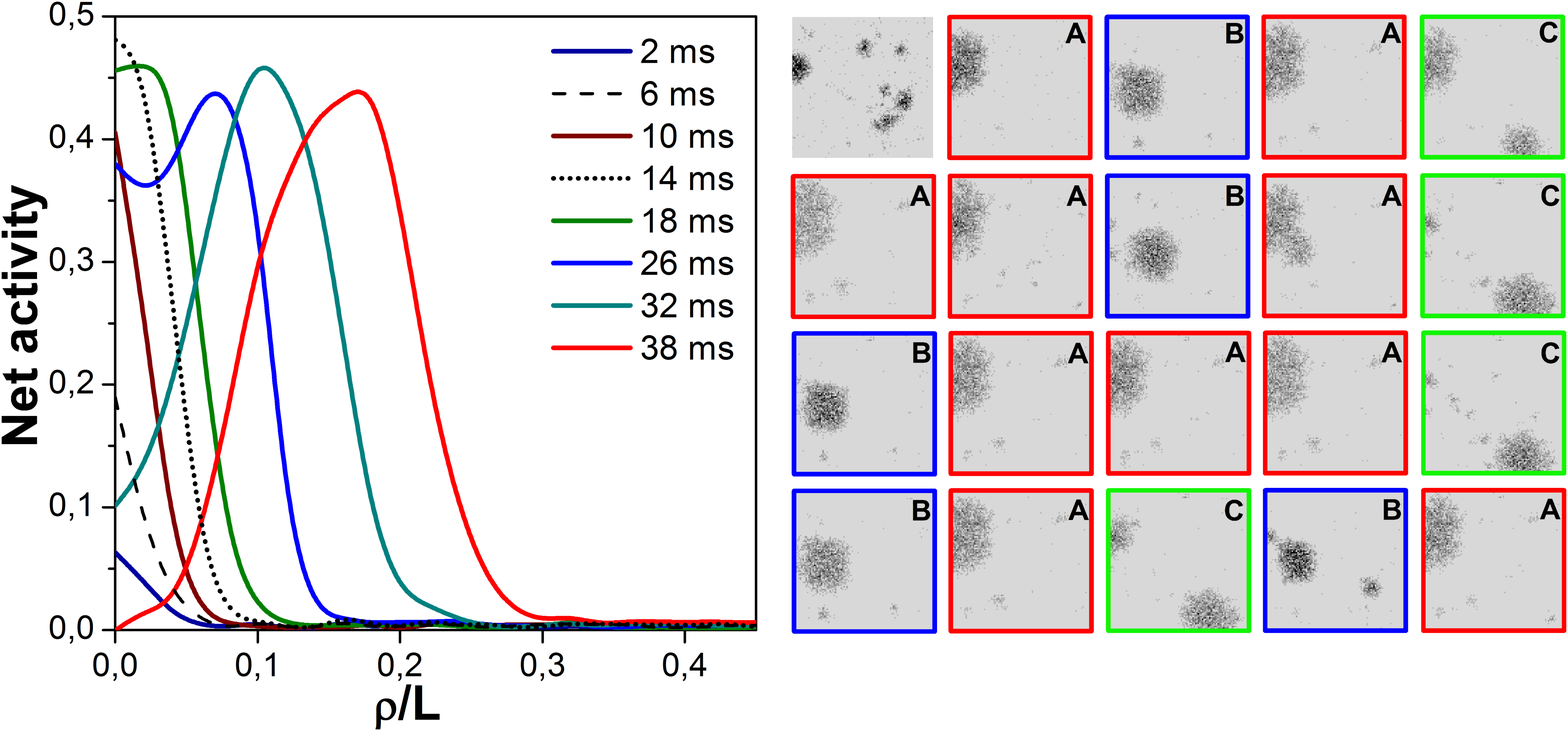}
\caption{Figure 5. Left graph: The dynamics of spiking activity front for the initial stage of the network spike shown in Fig. 4 (see frames \textbf{A} - \textbf{C}) with the relative times indication. The vertical axis denotes the spatially-averaged spiking activity of the network, the horizontal one shows the distance $\rho$ from the nucleation center. Right graph: Spatial locations of the nucleation centers of twenty subsequent network spikes occurred after blocking the inhibitory neurons for the same network as that in Fig. 4, where only the first fourteen of these network spikes are shown. Three nucleation centers (A, B, C) with different relative rates of network spike generation (10/20, 5/20 and 4/20, respectively) are clearly distinguishable. Black dots depict the spatial spiking activity of neurons during the first 40 ms after the network spike onset that was determined by the excess of a threshold value for the network spiking activity.}
\label{Fig5}
\end{figure*}

\begin{figure*}[!t]
\centering
\includegraphics[width=0.85\textwidth]{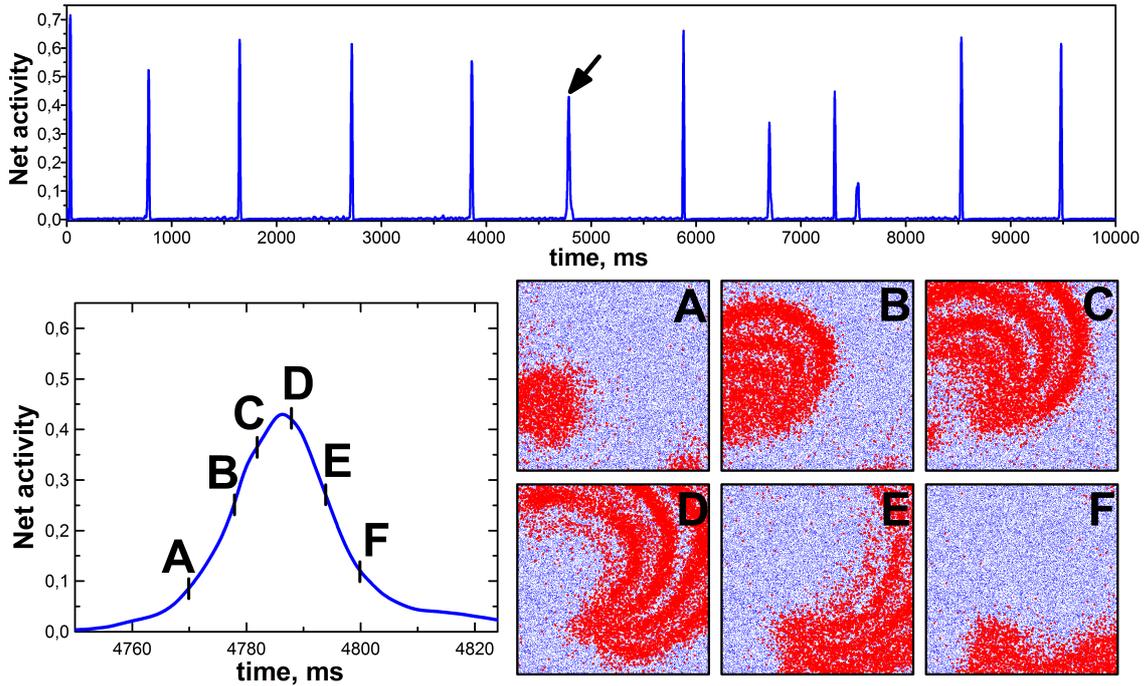}
\caption{Figure 6. Multiarmed spiral wave with the drifting center during the network spike (marked by the arrow) for the network of 50 thousand neurons at $\lambda = 0.04L$ that gives approx. 23 million interneuronal connections in the network, with approx. 460 outgoing connections per neuron. Other simulation parameters were taken the same as for the network in Fig. 4. Inhibitory neurons (20\% of the total) are not blocked during the whole time of the simulation. Blocking inhibitory neurons at $\lambda = 0.04L$ results in disappearance of the TUM-regime (see region III of the phase diagram in Fig. 3 A).}
\label{Fig6}
\end{figure*}

\bigskip

\textbf{4. Discussion}

\bigskip

Despite some theoretical advances \cite{25,38,39,40,41,42,43,44,45}, a theory for the origin of the nucleation centers of network spikes, enabling prediction of the number and locations of primary nucleation centers before carrying out the dynamic simulations, is currently absent. We have excluded the influence of fluctuations of the spatial distribution density of neurons by placing the neurons strictly periodically in the nodes of a square lattice - the nucleation centers still occurred (Fig. 7, upper panel). One-to-one correspondence was not observed between the locations of local maxima of spatial density of pacemaker neurons and primary nucleation centers. Nucleation centers occurred even at identical values of synaptic parameters (see \eqref{Isyn}, \eqref{xyz}) for all synapses of the network and also in the case where pacemaker neurons did not have any incoming connections. In addition, a redistribution (i.e., a new sampling) of the background currents during the simulation led to the change in the number of primary nucleation centers, their locations and relative rates of the network spike generation.

Modifications of the functional dependence of the probability of interneuronal connection formation on the distance between neurons (e.g., $p_{con}(r)=\theta(\lambda-r)$, where $\theta(\ldots)$ is unit step function), provided that (i) the average number of outgoing connections per neuron remains the same in the order of magnitude and (ii) neurons located far (compared with the characteristic distance $\lambda$, $L/\sqrt{N}<\lambda\ll L$) from each other practically do not form connections, also do not lead to the disappearance of nucleation centers. On the other hand, if the probability of interneuronal connection formation was independent of the distance between neurons (i.e., $p_{con}(r)=const$, given that the average number of outgoing connections per neuron remains unchanged), the nucleation centers did not arise - synchronization of spiking activity occurred spatially uniform throughout the network (lower panel in Fig. 7, a total of 5 such simulations were performed). Here, it may be significant that the variance of the total number of interneuronal connections of the network with $p_{con}(r)=e^{-r/\lambda}$, corresponding to the Bernoulli trials with variable probabilities of success, reaches its maximum at $p_{con}(r)=const$, i.e. in the limit of standard binomial distribution \cite{46}. Nevertheless, the results of simulations strongly suggest that the spatial proximity of the majority of network interneuronal connections (cp. \cite{42}) is important for the formation of nucleation centers. Moreover, the evaluation of network-averaged values of the shortest path length $\left\langle SPL\right\rangle$ and clustering coefficient $\left\langle C\right\rangle$ \cite{47} indicates that the neuronal networks exhibiting nucleation centers during network spikes belong to small-world networks (this correlates with findings \cite{48}). In particular, $\left\langle SPL\right\rangle\approx4.3$ and $\left\langle C\right\rangle\approx0.1$ for the network in Fig. 4, $\left\langle SPL\right\rangle\approx3.8$ and $\left\langle C\right\rangle\approx0.1$ for the network in Fig. 7 (upper panel), and $\left\langle SPL\right\rangle\approx3.4$ and $\left\langle C\right\rangle\approx10^{-3}$ for truly random (binomial) network in Fig. 7 (lower panel).
\begin{figure*}[!t]
\centering
\includegraphics[width=0.95\textwidth]{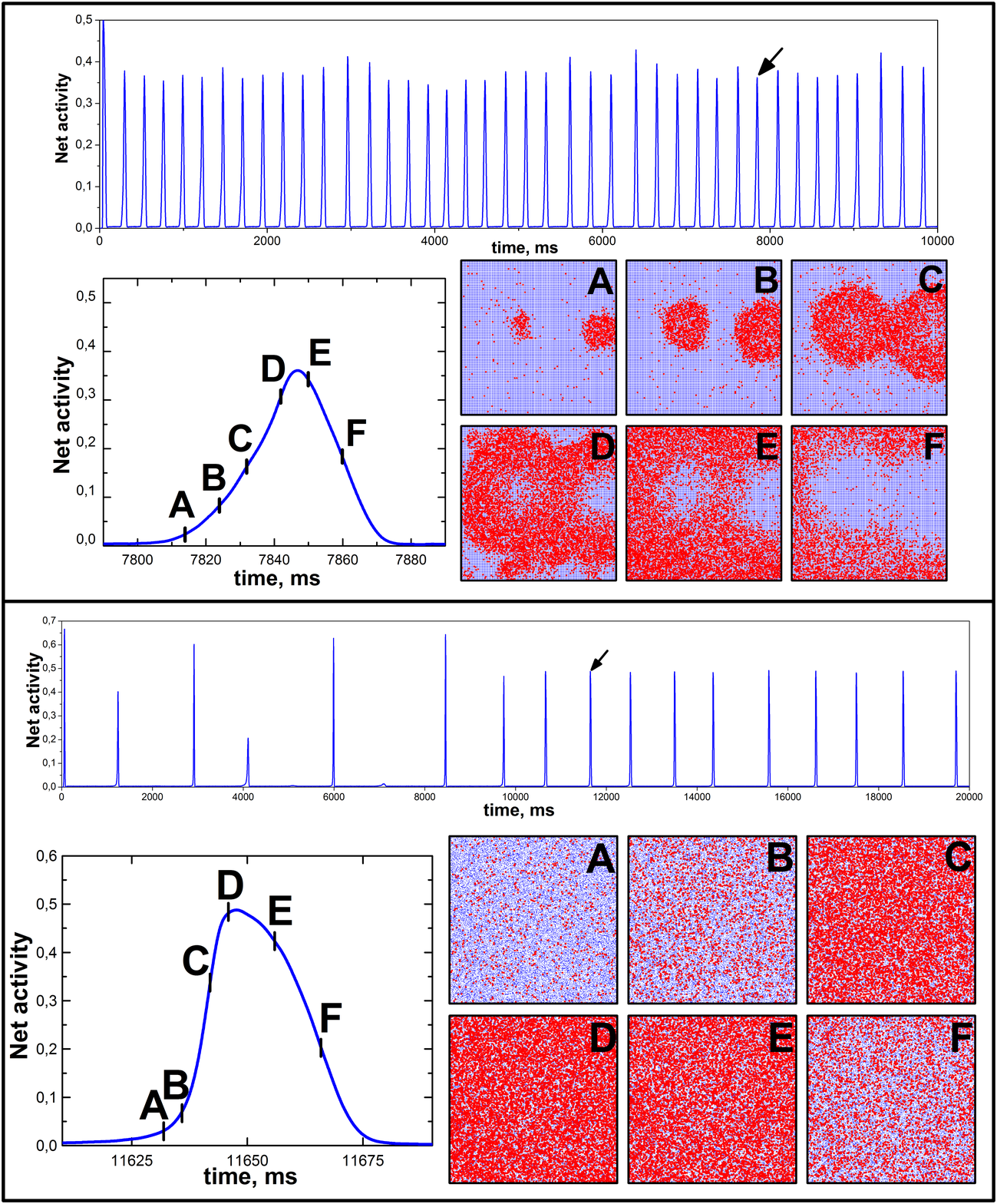}
\caption{Figure 7. Upper panel: Spatial dynamics of the network spike (marked by the arrow) for the network of 50625 neurons located in $225\times225$ nodes of a square lattice having period $a=0.004L$. Inhibitory neurons (20\% of the total) are blocked from the outset of the simulation. The dependence $p_{con}(r)$, with $\lambda = 0.014L$, and other simulation parameters were taken the same as for the network in Fig. 4. On frames \textbf{A} and \textbf{B} two nucleation centers are clearly visible. Lower panel: A similar simulation for the network of 50 thousand neurons at $p_{con}(r)=const=6.4\cdot10^{-4}$, with the same average number of outgoing connections per neuron and other simulation parameters as those for the network in Fig. 4. It is seen that there are no nucleation centers.}
\label{Fig7}
\end{figure*}

In general, our findings on nucleation centers, circular traveling waves and drifting spiral waves of spiking activity are in qualitative agreement with the already known. In particular, the nucleation centers and circular traveling waves during a network spike were directly observed experimentally in neuronal cultures \cite{4}, and the spiral waves were observed in disinhibited neocortical slices \cite{49,50,51}. The similar effects (e.g., circular and multiarmed spiral waves), regardless to the occurrence of network spikes, were also obtained in previous computational studies \cite{52,53,54,55,56} using different models of the neuronal network.

\bigskip

\textbf{5. Conclusion}

\bigskip

A relatively simple model of a planar neuronal culture is described that demonstrates in simulations the regime of repetitive network spikes emerging in a small number of spatial nucleation centers, the locations of which are unique for a given network implementation. In fact, the number and locations of primary nucleation centers are dynamic "marks of distinction" of neuronal cultures from each other.

More specifically, we have shown that (i) in spatially uniform networks of excitatory neurons, a typical network spike has complex spatial dynamics with a few nucleation centers, (ii) the spatial nucleation centers of a network spike appear if the majority of connections between neurons are the local ones that implies a small-world topology of the neuronal network, (iii) the nucleation centers are not nested in fluctuations of spatial density of neurons, and (iv) in the networks of excitatory and inhibitory neurons with relatively high density of connections a network spike can be a multiarmed spiral wave with the drifting center.

It is worth noting that transient spatial dynamics of a network spike in real neuronal cultures can be directly visualized with high spatial and temporal resolution using multi-transistor arrays \cite{57} or advanced standard microelectrode arrays \cite{58}, as well as using optical techniques such as calcium imaging \cite{4,5} or voltage-sensitive dye imaging \cite{51}. Therefore the results of simulations similar to those conducted in this study allow a direct comparison with experimental observations.

\end{document}

\begin{center}
Supplementary material to the article \\ \textbf{A spatially resolved network spike in model neuronal cultures reveals \\
nucleation centers, circular traveling waves and drifting spiral waves}
\end{center}
\underline{1. Derivation of the probability density $P(r)$ for a given distance $r$ between two random points}\\
\underline{inside a square of side $L$.}

\bigskip

1.1. Probability density of the random variable $\eta=(\xi_{1}-\xi_{2})^{2}$, given that random variables $\xi_{1}$ and $\xi_{2}$ are independent and uniformly distributed in the interval $[0, L]$. For $0\leq x\leq L^{2}$ the cumulative distribution function reads%
\[
F_{1}(x)=\text{Prob}\left[  (\xi_{1}-\xi_{2})^{2}\leq x\right]  =S/L^{2},
\]
where $S=x+2\sqrt{x}(L-\sqrt{x})$ is area of the figure determined by inequalities $\xi_{2}-\sqrt{x}<\xi_{1}<\xi_{2}+\sqrt{x}$ within the square on the coordinate plane $(\xi_{1},\xi_{2})$. The probability density%
\[
P_{1}(x)=\frac{dF_{1}}{dx}=\frac{1}{L}\left[  1/\sqrt{x}-1/L\right]
\theta(L^{2}-x)\theta(x),
\]
where $\theta(x)$ is unit step function: $\theta(x)=1$, if $x>0$, otherwise $\theta(x)=0$.

1.2. Probability density of the sum $\sigma=\eta_{1}+\eta_{2}=(\xi_{1}-\xi_{2})^{2}+(\xi_{3}-\xi_{4})^{2}$, given that all $\xi_{i}$ are independent and uniformly distributed in the interval $[0, L]$. The cumulative distribution function%
\[
F_{2}(y)=\text{Prob}\left[(\xi_{1}-\xi_{2})^{2}+(\xi_{3}-\xi_{4})^{2}\leq
y\right]  =
{\displaystyle\iint\limits_{x_{1}+x_{2}<y}}
P_{1}(x_{1})P_{1}(x_{2})dx_{1}dx_{2}=%
{\displaystyle\int\limits_{-\infty}^{+\infty}}
dx_{1}P_{1}(x_{1})%
{\displaystyle\int\limits_{-\infty}^{y-x_{1}}}
dx_{2}P_{1}(x_{2}).
\]
Accordingly, the probability density is given by
\[
P_{2}(y)=\frac{dF_{2}}{dy}=
{\displaystyle\int\limits_{-\infty}^{+\infty}}
dx_{1}P_{1}(y-x_{1})P_{1}(x_{1})=J_{1}(y)\theta(L^{2}-y)+J_{2}(y)\theta
(y-L^{2})\theta(2L^{2}-y).
\]
Here $J_{1}(y)$ and $J_{2}(y)$ stand for%
\begin{gather*}
J_{1}(y)={\int\limits_{0}^{y}}dx_{1}\frac{1}{L}\left[ \frac{1}{\sqrt{y-x_{1}}}-\frac{1}{L}\right] \frac{1}{L}\left[ \frac{1}{\sqrt{x_{1}}}-\frac{1}{L}\right] =\frac{1}{L^{2}}\left[ \pi -\frac{4}{L}\sqrt{y}+\frac{y}{L^{2}}\right], \\
J_{2}(y)={\int\limits_{y-L^{2}}^{L^{2}}}dx_{1}\frac{1}{L}\left[ \frac{1}{\sqrt{y-x_{1}}}-\frac{1}{L}\right] \frac{1}{L}\left[ \frac{1}{\sqrt{x_{1}}}-\frac{1}{L}\right] =\frac{2}{L^{2}}\left[ 2\arcsin (\frac{L}{\sqrt{y}})-\frac{\pi }{2}+\frac{2}{L}\sqrt{y-L^{2}}-\frac{y}{2L^{2}}-1\right].
\end{gather*}%
1.3. Finally, the probability density of a given distance $\rho=\sqrt{\sigma}$ between two random points inside a square of side $L$. The cumulative distribution function reads%
\[
F(r)=\text{Prob}\left[  \rho=\sqrt{(\xi_{1}-\xi_{2})^{2}+(\xi_{3}-\xi
_{4})^{2}}\leq r\right]  =
{\displaystyle\int\limits_{0}^{r^{2}}}
P_{2}(y)dy=
{\displaystyle\int\limits_{0}^{r}}
P_{2}(x^{2})2xdx\equiv
{\displaystyle\int\limits_{0}^{r}}
P(x)dx,
\]
so that the probability density $P(r)=P_{2}(r^{2})2r$. Explicitly, we get%
\[
P(r)=\begin{cases}
\frac{2r}{L^{2}}\left[\pi-\frac{4}{L}r+\frac{r^{2}}{L^{2}}\right], \text{  } r\leq L  \\
\frac{4r}{L^{2}}\left[2\arcsin(\frac{L}{r})-\frac{\pi}{2}%
+\frac{2}{L}\sqrt{r^{2}-L^{2}}-\frac{r^{2}}{2L^{2}}-1\right], \text{  } L<r\leq\sqrt{2}L
\end{cases}
\]
with ${\displaystyle\int\limits_{0}^{\sqrt{2}L}}P(r)dr=1$ and $\left\langle r\right\rangle ={\displaystyle\int\limits_{0}^{\sqrt{2}L}}rP(r)dr=\frac{2+\sqrt{2}+5\ln(1+\sqrt{2})}{15}L\approx0.52L$,
$\left\langle r^{2}\right\rangle ={\displaystyle\int\limits_{0}^{\sqrt{2}L}}r^{2}P(r)dr=\frac{1}{3}L^{2}$, $\sigma_{r}=\sqrt{\left\langle r^{2}\right\rangle -\left\langle r\right\rangle ^{2}}\approx0.25L$.

Note that $P(r)$ has maximum $\approx1.4/L$ at $r/L=(4-\sqrt{16-3\pi})/3\approx0.48$ and function $P(r)\exp(-r/\lambda)$ reaches its maximum at $r/L\approx b-(4/\pi)b^{2}$ given that $b=\lambda/L\lesssim0.1$. The average value of $\exp(-r/\lambda)$, evaluated below, determines $N_{con}(\lambda)$ in the article's main text.%
\begin{align*}
\left\langle \exp(-r/\lambda)\right\rangle ={\displaystyle\int\limits_{0}^{\sqrt{2}L}}\exp(-r/\lambda)P(r)dr=I_{1}(\lambda)+I_{2}(\lambda),\\
I_{1}(\lambda)={\displaystyle\int\limits_{0}^{L}}\exp(-r/\lambda)P(r)dr,\text{ \ \ }I_{2}(\lambda)={\displaystyle\int\limits_{L}^{\sqrt{2}L}}\exp(-r/\lambda)P(r)dr.
\end{align*}
A straightforward calculation of $I_{1}(\lambda)$ gives%
\[
I_{1}(\lambda)=2b^{2}\left(\pi-8b+6b^{2}\right)-2b\left(\left(\pi-3\right)-\left(5-\pi\right)b-2b^{2}+6b^{3}\right)\exp(-1/b),
\]%
where $b=\lambda/L$. As for $I_{2}(\lambda)$, at $b\gg1$ this can be approximated by%
\[
I_{2}(\lambda)\approx\left(19/6-\pi\right)\exp(-1/b).
\]
In spite of this approximation, the resulting analytical dependence $N_{con}(\lambda)$ is practically indistinguishable from that obtained by direct numerical integration for $0\leq b \leq 10$.
\bigskip
\bigskip
\pagebreak

\underline{2. Parameters used for the simulations shown in Fig. 4 of the article's main text.}

\bigskip

\textbf{2.1. LIF-neuron parameters}:

Relaxation time of the neuron's potential, $\tau _{m}=20$ ms.

Electrical (input) resistance of the neuron, $R_{m}=1$ G$\Omega$.

Resting potential, $V_{rest}=0$ mV.

Reset (after-spike) potential, $V_{reset}=13.5$ mV.

Threshold potential, $V_{th}=15$ mV.

Critical current for self-sufficient spiking, $I_{c}=(V_{th}-V_{rest})/R_{m}=15$ pA.

Refractory period, $\tau_{ref}=3$ ms for excitatory neurons, $\tau_{ref}=2$ ms for inhibitory neurons.

\textbf{2.2. Parameters for the background currents distribution}:

The distribution is non-negative normal distribution, with the mean $\mu=7.7$ pA and the standard deviation $\sigma=4.0$ pA, bounded above by $I_{\max}=20$ pA,
\[
P(I_{bg},\mu,\sigma)=\frac{1}{\sqrt{2\pi\sigma^{2}}}\exp(-\frac{(I_{bg}-\mu)^{2}}{2\sigma^{2}})\theta(I_{bg})\theta(I_{\max}-I_{bg}).
\]

The corresponding function {\ttfamily gauss()} written in C and used for generating the normal distributions is given by%
%\pagebreak
{\setstretch{1.0}
\begin{lstlisting}[language=C,backgroundcolor=\color{light-gray}]
double gauss (double mean, double sd, double min, double max){
int i;
double s;
while (1){
s = 0;
for (i=0; i< 12; i++){ s += ((double) rand() / (RAND_MAX)); }
s = s - 6;
if (((mean + s*sd) > min)&&((mean + s*sd) < max))
    return mean + s*sd;
}
}
\end{lstlisting}
}

\textbf{2.3. Synaptic parameters}:

All synaptic parameters, except $\tau _{I}$, were distributed normally using {\ttfamily gauss()} function with the mean values $\mu$ described below. Standard deviations for all distributed parameters equal $0.5\mu$, the maximal (or minimal, if the parameter is negative) values of the distributions equal $4\mu$ (or 1, if $4\mu>1$ for parameter $U$), and the minimal (maximal, if the parameter is negative) values equal zero (or time-step, for the time constants).

Maximal amplitudes of synaptic currents: $A_{ee}=38$ pA, $A_{ei}=54$ pA, $A_{ie}=A_{ii}=-72$ pA. The first lowercase index here denotes the type ($e$ = excitatory, $i$ = inhibitory) of presynaptic neuron and the second index stands for the type of postsynaptic neuron.

Recovery time constants: $\tau _{rec, ee}=\tau _{rec, ei}=800$ ms, $\tau _{rec, ie}=\tau _{rec, ii}=100$ ms.

Decay time constant of postsynaptic currents, $\tau _{I}=3$ ms.

Parameter $U$ values: $U_{ee}=U_{ei}=0.5$, $U_{ie}=U_{ii}=0.04$.

Time constants for synaptic strength facilitation: $\tau_{facil, ie}=\tau_{facil, ii}=1000$ ms.

\textbf{2.4. Spatial and network-topology parameters}:

Point-like neurons were uniformly distributed over the square area $L\times L$, with $L=1$ mm by default, so that the resulting network was also spatially uniform. All spatial variables are measured in units of $L$. The probability of making a synaptic connection decreases exponentially as a function of the relative distance $r$ between the neurons, $p_{con}(r)=e^{-r/\lambda}$, where $\lambda=0.01L$.

Temporal delays resulting from the propagation of spikes along the axons (modeled by straight lines) were calculated as $\tau_{del}=\tau_{del, min}+r/v_{sp}$, where $\tau_{del, min}=0.2$ ms is the minimal axonal delay, $r$ is the distance (in mm) between the neurons, or the axon length, and $v_ {sp} = 0.2$ mm/ms is the constant speed of spike propagation along the axon.

%\pagebreak
\textbf{2.5. Simulation parameters}:

Simulation time step, $dt=0.1$ ms.

Averaging time for the network spiking activity graphs, $t_{avg}=2$ ms.

Total simulation time, $t_{sim}=2\cdot10^{4}$ ms.

In order to elucidate the role of inhibitory neurons (20\% of total neuron number in the network), by default we used the protocol where, for the same network, inhibitory neurons were blocked (i.e., their membrane potentials were clamped to the resting potential $V_{rest}$) during the second half of the simulation time.

Finally, the initial conditions for the dynamic variables were the same for all neurons, $V(t=0)=V_{rest}$, and for all synapses: $x(t=0)=0.98$, $y(t=0)=z(t=0)=0.01$.

\bigskip

All numerical simulations of neural-network spiking dynamics were performed on the custom-made neurosimulator \textit{NeuroSim-TM} written in C and the visualization of spatio-temporal patterns of the network spiking activity was made by the custom-made visualization software \textit{Spatial Activity Monitor} written in C++ using the freely distributable libraries of the \textit{Qt} framework. The simulations were performed on the workstation equipped with quad-core Intel Core i7 processor and 16 GB of RAM. A typical simulation of 20 seconds of spiking activity for a network consisting of 50 thousand neurons, with on average 32 outgoing connections per neuron, took about $35\pm2$ hours.